\documentclass[12pt]{article}
\newcommand{\text}{\rm}

\begin{document}

\title{\textbf{The Diagonal Ghost Equation Ward Identity for Yang-Mills Theories in
the Maximal Abelian Gauge}}
\author{A.R. Fazio$^{\mathrm{(a)}}$, V.E.R. Lemes$^{\,\mathrm{(b)}}$, M.S. Sarandy$%
^{\,\mathrm{(c)}}$ \and and S.P. Sorella$^{\,\mathrm{(b)}},$\textbf{\ }%
\vspace{4mm} \\
%EndAName
{\small {$^{\mathrm{(a)}}$\textit{Dipartimento di Fisica, Universit\'{a}
Statale di Milano {and\thinspace INFN},}}} {\small {\textit{\ }}}\\
{\small {\textit{via Celoria 16, I-20133, Milano, Italy.}}}\vspace{2mm}\\
{\small {$^{\mathrm{(b)}}$\textit{\ UERJ, Universidade do Estado do Rio de
Janeiro}}} \\
{\small {\textit{\ Rua S\~{a}o Francisco Xavier 524, 20550-013 Maracan\~{a},
Rio de Janeiro, Brazil.}}}\vspace{2mm}\\
{\small {$^{\mathrm{(c)}}\,$\textit{CBPF, Centro Brasileiro de Pesquisas
F\'{\i}sicas}}}\\
{\small \textit{\ Rua Xavier Sigaud 150, 22290-180 Urca, Rio de Janeiro,
Brazil.}}\vspace{2mm}}
\maketitle

\begin{abstract}
A BRST perturbative analysis of $SU(N)$ Yang-Mills theory in a class of
maximal Abelian gauges is presented. We point out the existence of a new
nonintegrated renormalizable Ward identity which allows to control the
dependence of the theory from the diagonal ghosts. This identity, called the
diagonal ghost equation, plays a crucial role for the stability of the model
under radiative corrections implying, in particular, the vanishing of the
anomalous dimension of the diagonal ghosts. Moreover, the Ward identity
corresponding to the Abelian Cartan subgroup is easily derived from the
diagonal ghost equation. Finally, a simple proof of the fact that the beta
function of the gauge coupling can be obtained from the vacuum polarization
tensor with diagonal gauge fields as external legs is given. A possible
mechanism for the decoupling of the diagonal ghosts at low energy is also
suggested.
\end{abstract}

\vfill\newpage\ \makeatother

\renewcommand{\theequation}{\thesection.\arabic{equation}}

\section{Introduction}

The understanding of the color confinement has been a challenging issue in
theoretical physics since long time, being at present subject of intensive
research. Certainly, the idea that confinement could be interpreted as a
dual Meissner effect for type II superconductors \cite{mand,thooft} is both
attractive and very promising.

An important ingredient in order to implement this program is the mechanism
of the Abelian projection introduced by 't Hooft \cite{thooft}, which
consists of reducing the gauge group to an Abelian subgroup, commonly
identified with the Cartan subgroup, by means of a partial gauge fixing.
This procedure starts by decomposing the gauge field into its diagonal and
off-diagonal parts. The diagonal components correspond to the generators of
the Cartan subgroup and behave as photons. The off-diagonal components are
charged with respect to the Abelian residual subgroup and can become
massive, being not protected by gauge invariance anymore. The appearance of
this mass scale allows for the decoupling of the off-diagonal fields at low
energy. Moreover, the Abelian projected theory turns out to contain magnetic
monopoles, whose condensation should account for the confinement of all
chromoelectric charges.

Lattice calculations \cite{kronf,susuki} have provided evidences for the
Abelian dominance hypothesis, according to which QCD\ in the low energy
regime is described by an effective Abelian theory. This supports the
realization of confinement through a dual Meissner effect, although the
infrared Abelian dominance in lattice simulations seems not to be a general
feature of any Abelian gauge \cite{digiacomo}.

Furthermore, many conceptual points remain to be clarified in order to
achieve an analytic derivation of the Abelian dominance directly from the
QCD Lagrangian. One crucial question is to understand how the off-diagonal
fields acquire mass, so that they can decouple at low energy.

Recently, the authors \cite{kondo,schaden} have proposed an original
mechanism for generating a mass term for the off-diagonal fields. They make
use of the maximal Abelian gauge (MAG) which amounts to choose a nonlinear
gauge condition for the off-diagonal components of the gauge field.
Therefore, the introduction of a four-ghost self-interaction term \cite{mlp}
is required for renormalizability. As discussed in \cite{kondo,schaden}, the
four-ghost term is responsible for the existence of a nonperturbative
ghost-antighost condensation which provides a dynamical mass generation for
all off-diagonal gauge and ghost fields. It is worth underlining that this
mechanism shares great analogy with the BCS\ gap equation for
superconductivity \cite{bcs}. Another peculiar aspect of the gauge fixing
adopted in \cite{kondo,schaden} is the presence of a unique gauge parameter
instead of the usual pair of independent parameters associated respectively
to the MAG gauge fixing condition and to the four-ghost self-interaction.
This particular choice of the gauge parameters allows for both BRST and
anti-BRST invariances, which may provide a possible consistent
interpretation of the ghost-antighost condensation in terms of a spontaneous
breakdown of a global $SL(2,R)$ symmetry present in the theory \cite{schaden}%
.

The aim of this work is to point out some new properties of the gauge
proposed in \cite{kondo,schaden}. We shall be able to prove indeed that, as
a consequence of having a unique gauge parameter, a new Ward identity
arises, which allows to control the dependence of the theory from the
diagonal ghosts. A remarkable feature of this Ward identity, which we shall
call the \textit{diagonal ghost equation}, relies on the fact that it holds
at the nonintegrated level, a property which will have far reaching
consequences. For instance, the Abelian Ward identity corresponding to the
Cartan subgroup follows by commuting the Slavnov-Taylor identity with the
diagonal ghost equation. Furthermore, this equation is easily extended at
the quantum level implying the stability under radiative corrections of the
gauge fixing condition. The diagonal ghost equation imposes also strong
constraints on the Callan-Symanzik equation of the theory. In particular,
the anomalous dimension of the diagonal ghosts turns out to vanish to all
orders of perturbation theory and a simple proof of the fact that the beta
function of the gauge coupling can be computed directly from the two-point
Green function with diagonal gauge fields as external legs can be given.
According to \cite{quandt}, this result is interpreted as an evidence for
the Abelian dominance.

The paper is organized as follows. In section 2\ we discuss the case of $%
SU(2)$\ Yang-Mills in the MAG, deriving the diagonal ghost equation Ward
identity and its consequences at the quantum level. Section 3 is devoted to
the extension of all results to the general case of $SU(N)$. In Section 4 we
present the conclusions.

\section{SU(2) Yang-Mills gauge theory in the MAG}

\subsection{Classical aspects}

Let $\mathcal{A}_\mu $ be the Lie algebra valued connection for the gauge
group $SU(2),$ whose generators $T^A\,\;(A=1,..,3)\,$

\begin{equation}
\left[ T^A,T^B\right] =\varepsilon ^{ABC}T^C\,\,  \label{la}
\end{equation}
are chosen to be antihermitean and to obey the orthonormality condition $\,%
\mathrm{Tr}\left( T^AT^B\right) =\delta ^{AB}$. Following \cite{thooft}, we
decompose the gauge field into its off-diagonal and diagonal parts

\begin{equation}
\mathcal{A}_\mu =\mathcal{A}_\mu ^AT^A=A_\mu ^aT^a+A_\mu T^{\,3},  \label{cd}
\end{equation}
where $a=1,2$ and $T^3$ is the diagonal generator of the Cartan subgroup.
Analogously, decomposing the field strength we obtain

\begin{equation}
\mathcal{F}_{\mu \nu }=\mathcal{F}_{\mu \nu }^AT^A=F_{\mu \nu }^aT^a+F_{\mu
\nu }T^{\,3},  \label{fs}
\end{equation}
with the off-diagonal and diagonal parts given respectively by

\begin{eqnarray}
F_{\mu \nu }^a &=&D_\mu ^{ab}A_\nu ^b-D_\nu ^{ab}A_\mu ^b\,,  \nonumber \\
F_{\mu \nu } &=&\partial _\mu A_\nu -\partial _\nu A_\mu +\varepsilon
^{ab}A_\mu ^aA_\nu ^b\,,  \label{fscomp}
\end{eqnarray}
where the covariant derivative $D_\mu ^{ab}$ is defined with respect to the
diagonal component $A_\mu $

\begin{equation}
D_\mu ^{ab}\equiv \partial _\mu \delta ^{ab}-\varepsilon ^{ab}A_\mu
\,\,\,\,\,\,,\,\,\,\varepsilon ^{ab}\equiv \varepsilon ^{ab3}\,\,\,.
\label{cder}
\end{equation}
Rewriting the Yang-Mills action in terms of the off-diagonal and diagonal
fields we get

\begin{equation}
S_{\mathrm{YM}}=-\frac 1{4g^2}\int d^4x\,\left( F_{\mu \nu }^aF^{a\mu \nu
}+F_{\mu \nu }F^{\mu \nu }\right) \;.  \label{sym}
\end{equation}
In order to quantize the theory, we shall adopt the so called MAG condition 
\cite{kronf}, which amounts to fix the value of the covariant derivative $%
(D_\mu ^{ab}A^{b\mu })$ of the off-diagonal components. However, this
condition being nonlinear, a four-ghost self-interaction term is needed to
guarantee the perturbative renormalizability \cite{mlp}. Therefore, the
corresponding gauge fixing term is found to be

\begin{equation}
S_{\mathrm{MAG}}=s\,\int d^4x\,\left( \overline{c}^a\left( D_\mu
^{ab}A^{b\mu }+\frac \alpha 2b^a\right) -\frac \xi 2\varepsilon ^{ab}%
\overline{c}^a\overline{c}^bc\right)  \label{smag}
\end{equation}
where $\alpha $ and $\xi $ are gauge parameters and the BRST operator $s$
acts as

\begin{eqnarray}
&& 
\begin{tabular}[t]{ll}
$sA_\mu ^a=-\left( D_\mu ^{ab}c^b+\varepsilon ^{ab}A_\mu ^bc\right) ,$ & $%
\,sA_\mu =-\left( \partial _\mu c+\varepsilon ^{ab}A_\mu ^ac^b\right) ,$ \\ 
$sc^a=\varepsilon ^{ab}c^bc\,,$ & $\,\,sc=\frac 12\varepsilon ^{ab}c^ac^b,$
\\ 
$s\overline{c}^a=b^a\,,$ & $\,sb^a=0\;.$%
\end{tabular}
\nonumber  \label{BRST} \\
&&  \label{BRST}
\end{eqnarray}
Here $c^a,c$ denote the off-diagonal and the diagonal components of the
Faddeev-Popov ghost field, while $\overline{c}^a,b^a$ are the off-diagonal
antighost and Lagrange multiplier. We also observe that the BRST\
transformations $\left( \ref{BRST}\right) $ have been obtained by their
standard form upon projection on the off-diagonal and diagonal components of
the fields. Expression $\left( \ref{smag}\right) $ is easily worked out and
yields

\begin{eqnarray}
S_{\mathrm{MAG}} &=&\int d^4x\left( b^a\left( D_\mu ^{ab}A^{b\mu }+\frac
\alpha 2b^a\right) +\overline{c}^aD_\mu ^{ab}D^{\mu bc}c^c+\overline{c}%
^a\varepsilon ^{ab}\left( D_\mu ^{bc}A^{c\mu }\right) c\right.  \nonumber \\
&&\left. \,\,\,\,\,\,\,\,\,\,\,\,\,\,\,\,\,\,-\varepsilon ^{ab}\varepsilon
^{cd}\overline{c}^ac^dA_\mu ^bA^{c\mu }-\xi \varepsilon ^{ab}b^a\overline{c}%
^bc-\frac \xi 4\varepsilon ^{ab}\varepsilon ^{cd}\overline{c}^a\overline{c}%
^bc^cc^d\right) \;.  \label{smag2}
\end{eqnarray}
It should be remarked that the use of the MAG condition allows for the
existence of a residual local $U(1)$ invariance with respect to the diagonal
subgroup, which has to be fixed by imposing a suitable further gauge
condition on the diagonal component $A_\mu $ of the gauge field. Adopting
without loss of generality a linear Landau condition, the remaining gauge
fixing term is given by 
\begin{eqnarray}
S_{\mathrm{diag}} &=&s\,\,\int d^4x\,\,\overline{c}\partial ^\mu A_\mu \; 
\nonumber  \label{su1} \\
&=&\int d^4x\left( b\partial ^\mu A_\mu +\overline{c}\partial
^2c+\varepsilon ^{ab}\overline{c}\partial ^\mu \left( A_\mu ^ac^b\right)
\right) \;,  \label{su1}
\end{eqnarray}
where $\,\,\overline{c},b$ are the diagonal antighost and Lagrange
multiplier fields which transform as

\begin{equation}
s\overline{c}=b\,\,,\,\,sb=0\;.  \label{BRST2}
\end{equation}
In order to write down the Slavnov-Taylor identity, we follow the standard
procedure \cite{book} and introduce a set of \ invariant external classical
sources $(A_\mu ^{a*},A_\mu ^{*},c^{a*},c^{*})$ coupled to the nonlinear
BRST\ transformations, \textit{i.e.}

\begin{equation}
S_{\mathrm{ext}}=\int d^4x\left( A_\mu ^{a*}sA^{a\mu }+A_\mu ^{*}sA^\mu
+c^{a*}sc^a+c^{*}sc\right) .  \label{sext}
\end{equation}
Therefore, the complete action

\begin{equation}
\Sigma =S_{\mathrm{YM}}+S_{\mathrm{MAG}}+S_{\mathrm{diag}}+S_{\mathrm{ext}}
\label{ca}
\end{equation}
turns out to obey the Slavnov-Taylor identity

\begin{equation}
\mathcal{S}(\Sigma )=0\;,  \label{st}
\end{equation}
with

\begin{equation}
\mathcal{S}(\Sigma )=\int d^4x\left( \frac{\delta \Sigma }{\delta A_\mu ^{a*}%
}\frac{\delta \Sigma }{\delta A^{a\mu }}+\frac{\delta \Sigma }{\delta A_\mu
^{*}}\frac{\delta \Sigma }{\delta A^\mu }+\frac{\delta \Sigma }{\delta c^{a*}%
}\frac{\delta \Sigma }{\delta c^a}+\frac{\delta \Sigma }{\delta c^{*}}\frac{%
\delta \Sigma }{\delta c}+b^a\frac{\delta \Sigma }{\delta \overline{c}^a}+b%
\frac{\delta \Sigma }{\delta \overline{c}}\right) \;.  \label{st-exp}
\end{equation}
Let us also introduce for further use the linearized Slavnov-Taylor operator 
\cite{book}

\begin{eqnarray}
\mathcal{B}_\Sigma &=&\int d^4x\left( \frac{\delta \Sigma }{\delta A_\mu
^{a*}}\frac \delta {\delta A^{a\mu }}+\frac{\delta \Sigma }{\delta A^{a\mu }}%
\frac \delta {\delta A_\mu ^{a*}}+\frac{\delta \Sigma }{\delta A_\mu ^{*}}%
\frac \delta {\delta A^\mu }+\frac{\delta \Sigma }{\delta A^\mu }\frac
\delta {\delta A_\mu ^{*}}\right.  \nonumber  \label{lst} \\
&&\left. +\frac{\delta \Sigma }{\delta c^{a*}}\frac \delta {\delta c^a}+%
\frac{\delta \Sigma }{\delta c^a}\frac \delta {\delta c^{a*}}+\frac{\delta
\Sigma }{\delta c^{*}}\frac \delta {\delta c}+\frac{\delta \Sigma }{\delta c}%
\frac \delta {\delta c^{*}}+b^a\frac \delta {\delta \overline{c}^a}+b\frac
\delta {\delta \overline{c}}\right) ,  \label{lst}
\end{eqnarray}
which enjoys the property of being nilpotent

\begin{equation}
\mathcal{B}_\Sigma \mathcal{B}_\Sigma =0\;.  \label{nilp}
\end{equation}
We remark that, due to the choice of a linear condition for the diagonal
gauge fixing term $\left( \ref{su1}\right) ,$ the complete action $\Sigma $
obeys the additional constraint corresponding to the equation of motion of
the diagonal Lagrange multiplier $b$ \cite{book} 
\begin{equation}
\frac{\delta \Sigma }{\delta b}=\partial ^\mu A_\mu \;.  \label{gfc}
\end{equation}
As usual, commuting eq.$\left( \ref{gfc}\right) $ with the Slavnov-Taylor
identity $\left( \ref{st}\right) $ one gets

\begin{equation}
\frac{\delta \Sigma }{\delta \overline{c}}+\partial ^\mu \frac{\delta \Sigma 
}{\delta A^{*\mu }}=0\;,  \label{age}
\end{equation}
which implies that $\Sigma $ depends on the diagonal antighost $\overline{c}$
only through the combination $(A^{*\mu }+\partial ^\mu \overline{c}).$ Let
us end this subsection by displaying the quantum numbers of the fields and
external sources.

\begin{eqnarray*}
&& 
\begin{tabular}{|l|l|l|l|l|l|l|l|l|l|l|l|l|}
\hline
Field & $A_\mu ^a$ & $A_\mu $ & $c^a$ & $c$ & $\overline{c}^a$ & $\overline{c%
}$ & $b^a$ & $b$ & $A_\mu ^{a*}$ & $A_\mu ^{*}$ & $c^{a*}$ & $c^{*}$ \\ 
\hline
Gh. number & \thinspace 0 & \thinspace \thinspace 0 & 1 & \thinspace 1 & -1
& -1 & 0 & 0 & -1 & -1 & \thinspace -2 & -2 \\ \hline
Dimension & \thinspace 1 & $\,\,$1 & 0 & $\,$0 & $\,\,$2 & $\,\,$2 & $\,$2 & 
2 & \thinspace 3 & \thinspace \thinspace 3 & \thinspace \thinspace
\thinspace 4 & \thinspace 4 \\ \hline
\end{tabular}
\\
&\,\,\,\,&_{\,\,\,\,\,\,\,\,\,\,\,\,\,\,\,\,\,\,\,\,\,\,\,\,\,\,\,\,\,\,\,\,%
\,\,\,\,\mathrm{Table\,1.\,Ghost\,number\,\,and\,\,canonical\,\,dimension\,%
\,of\,\,the\,\,fields.\,}}
\end{eqnarray*}

\subsection{The diagonal ghost equation}

As one can see from the expression $\left( \ref{smag}\right) ,$ the gauge
fixing term corresponding to the MAG condition contains two independent
gauge parameters $\alpha ,\xi .$ It is known that the number of such
parameters can be reduced by requiring that the quantized action possesses
additional invariances as, for instance, the anti-BRST\ symmetry. This is
the case considered in \cite{kondo,schaden}, where the gauge parameter $%
\alpha $ is set to be equal to $\xi ,$ namely $\alpha =\xi $. However, it is
a remarkable fact that, in addition to the anti-BRST\ symmetry, this choice
allows for the existence of a further new Ward identity \ which accounts for
the dependence of the model from the diagonal ghost $c$. In order to
establish the form of the diagonal ghost equation, let us act with the
functional operator

\begin{equation}
\mathcal{G=}\frac \delta {\delta c}+\varepsilon ^{ab}\overline{c}^a\frac
\delta {\delta b^b}\;\;  \label{gop}
\end{equation}
on the complete action $\Sigma $ of eq.$\left( \ref{ca}\right) $. After a
straightforward algebra, we obtain

\begin{equation}
\frac{\delta \Sigma }{\delta c}+\varepsilon ^{ab}\overline{c}^a\frac{\delta
\Sigma }{\delta b^b}=-\partial ^2\,\overline{c}-\partial ^\mu A_\mu
^{*}+\varepsilon ^{ab}A_\mu ^{a*}A^{b\mu }-\varepsilon ^{ab}c^{a*}c^b+(\xi
-\alpha )\varepsilon ^{ab}b^a\overline{c}^b\;.  \label{qb}
\end{equation}
We see therefore that, besides terms which are purely linear in the quantum
fields, the right hand side of the eq.$\left( \ref{qb}\right) $ contains the
term $\varepsilon ^{ab}b^a\overline{c}^b$ which, being quadratic in the
quantum fields, is a composite operator which has to be properly defined and
renormalized at the quantum level. However, as it is apparent from $\left( 
\ref{qb}\right) $, the coefficient of the quadratic breaking $\varepsilon
^{ab}b^a\overline{c}^b$ happens to be precisely the difference $(\xi -\alpha
)$ between the two gauge parameters. Therefore, choosing $\alpha =\xi ,$ the
right hand side of the equation $\left( \ref{qb}\right) $ turns out to
contain breakings which are only linear in the quantum fields. Equation $%
\left( \ref{qb}\right) $ acquires thus the meaning of a Ward identity,
according to the Quantum Action Principle \cite{book}. In summary, setting $%
\alpha =\xi ,$ the complete action $\Sigma $ is constrained by the following
diagonal ghost equation

\begin{equation}
\mathcal{G}\Sigma =\Delta _{\mathrm{cl}}\;,  \label{ge}
\end{equation}
where $\Delta _{\mathrm{cl}}$ is given by

\begin{equation}
\Delta _{\mathrm{cl}}=-\partial ^2\,\overline{c}-\partial ^\mu A_\mu
^{*}+\varepsilon ^{ab}A_\mu ^{a*}A^{b\mu }-\varepsilon ^{ab}c^{a*}c^b\;.
\label{clbr}
\end{equation}
As already underlined, the above expression is linear in the quantum fields,
so that $\Delta _{\mathrm{cl}}$ is a classical breaking not affected by the
quantum corrections \cite{book}. Of course, the equality $\alpha =\xi $ is
here understood and will be assumed from now on.

Although being self-explanatory, it is worth observing that the Ward
identity $\left( \ref{qb}\right) $ allows in fact to control the dependence
of the theory from the diagonal ghost $c$. An important feature of this
equation is that it holds at the nonintegrated level, a property which will
imply many consequences. The first one can be obtained by acting with the
diagonal ghost operator $\left( \ref{gop}\right) $ on the Slavnov-Taylor
identity $\left( \ref{st}\right) ,$ yielding a further local Ward identity,
namely

\begin{equation}
\mathcal{GS}(\Sigma )=0\,\,\,\,\,\,\,\Rightarrow \;\;\mathcal{W}^{\mathrm{%
U(1)}}\Sigma =-\partial ^2b\;,  \label{gh-u1}
\end{equation}
where $\mathcal{W}^{\mathrm{U(1)}}\;$is the Ward operator

\begin{equation}
\mathcal{W}^{\mathrm{U(1)}}=\partial _\mu \frac \delta {\delta A_\mu
}+\varepsilon ^{ab}\left( A_\mu ^a\frac \delta {\delta A_\mu ^b}+c^a\frac
\delta {\delta c^b}+b^a\frac \delta {\delta b^b}+\overline{c}^a\frac \delta
{\delta \overline{c}^b}+A_\mu ^{a*}\frac \delta {\delta A_\mu
^{b*}}+c^{a*}\frac \delta {\delta c^{b*}}\right) \;.  \label{wop}
\end{equation}
The identity

\begin{equation}
\mathcal{W}^{\mathrm{U(1)}}\Sigma =-\partial ^2b\;,  \label{u1ident}
\end{equation}
is recognized to be the local $U(1)$ Ward identity corresponding to the
Cartan subgroup of $SU(2).\;$As in ordinary $QED,$ one sees that the
diagonal component $A_\mu $ of the gauge field behaves as a photon, while
all off-diagonal components play the role of charged matter fields. Observe
also that the classical linear breaking $\partial ^2b$ in the right hand
side of eq.$\left( \ref{u1ident}\right) $ stems from the diagonal gauge
fixing term $\left( \ref{su1}\right) $, in perfect analogy with $QED.$

We are now ready to analyse the quantum aspects. This will be the task of
the next subsection.

\subsection{Quantum aspects: renormalizability and invariant counterterm}

Let us begin by observing that all classical Ward identities, namely eqs.$%
\left( \ref{st}\right) $, $\left( \ref{ge}\right) $, $\left( \ref{u1ident}%
\right) $ and constraints $\left( \ref{gfc}\right) ,\left( \ref{age}\right) $
can be extended to all orders of perturbation theory without anomalies. This
result can be proven by following the algebraic set up presented in \cite
{book} and by making use of the general results on the BRST\ cohomology of
gauge theories \cite{mh}. It can be easily understood by observing that, in
the present case, the theory can be regularized in a gauge invariant way by
means of the dimensional regularization.

In order to determine the most general invariant counterterm which can be
freely added to all orders of perturbation theory, we perturb the classical
action $\Sigma $ by adding an arbitrary integrated local polynomial $\Sigma
^{\mathrm{c}}$ in the fields and external sources of dimension bounded by
four and with zero ghost number, and we require that the perturbed action $%
(\Sigma +\epsilon \Sigma ^{\mathrm{c}})$ satisfies the same Ward identities
and constraints as $\Sigma $ to the first order in the perturbation
parameter $\epsilon ,$ \textit{i.e.}

\begin{eqnarray}
\mathcal{S}(\Sigma +\epsilon \Sigma ^{\mathrm{c}}) &=&0+O(\epsilon ^2)\;, 
\nonumber  \label{con-1} \\
\mathcal{G}(\Sigma +\epsilon \Sigma ^{\mathrm{c}}) &=&\Delta _{\mathrm{cl}%
}\;+O(\epsilon ^2)\;,  \nonumber \\
\mathcal{W}^{\mathrm{U(1)}}(\Sigma +\epsilon \Sigma ^{\mathrm{c}})
&=&-\partial ^2b\,+O(\epsilon ^2)\;,  \nonumber \\
\frac{\delta (\Sigma +\epsilon \Sigma ^{\mathrm{c}})}{\delta b} &=&\partial
^\mu A_\mu +O(\epsilon ^2)\;,  \nonumber \\
\left( \frac \delta {\delta \overline{c}}+\partial ^\mu \frac \delta {\delta
A^{*\mu }}\right) (\Sigma +\epsilon \Sigma ^{\mathrm{c}}) &=&0\;+O(\epsilon
^2)\;,  \label{cont-1}
\end{eqnarray}
which amount to impose the following conditions on $\Sigma ^{\mathrm{c}}$

\begin{equation}
\mathcal{B}_\Sigma \Sigma ^{\mathrm{c}}=0\,,  \label{b-c}
\end{equation}

\begin{equation}
\mathcal{G}\Sigma ^{\mathrm{c}}=0\;,  \label{g-c}
\end{equation}

\begin{equation}
\mathcal{W}^{\mathrm{U(1)}}\Sigma ^{\mathrm{c}}=0\;,  \label{u1-c}
\end{equation}
and

\begin{equation}
\frac{\delta \Sigma ^{\mathrm{c}}}{\delta b}=0\;,\;\;\;\;\;\;\;\;\;\;\;\frac{%
\delta \Sigma ^{\mathrm{c}}}{\delta \overline{c}}+\partial ^\mu \frac{\delta
\Sigma ^{\mathrm{c}}}{\delta A^{*\mu }}=0\;.  \label{cont-two}
\end{equation}
Equations $\left( \ref{cont-two}\right) $ imply that $\Sigma ^{\mathrm{c}}$
is independent from the Lagrange multiplier $b$ and that the antighost $%
\overline{c}$ enters only through the combination $(A^{*\mu }+\partial ^\mu 
\overline{c}).$ Furthermore, from the general results on the cohomology of
Yang-Mills theories \cite{mh,book}, the most general solution of $\left( \ref
{b-c}\right) $ is found to be 
\begin{equation}
\Sigma ^{\mathrm{c}}=-\frac \sigma {4g^2}\int d^4x\,\left( F_{\mu \nu
}^aF^{a\mu \nu }+F_{\mu \nu }F^{\mu \nu }\right) +\mathcal{B}_\Sigma
\,\Delta ^{-1}\;,  \label{count}
\end{equation}
where $\sigma $ is an arbitrary parameter and $\Delta ^{-1}$ given by

\begin{eqnarray}
\Delta ^{-1} &=&\int d^4x\left( a_1\,c^{a*}c^a+\;a_2\,A_\mu ^{a*}A^{a\mu
}+a_3\,\xi \overline{c}^ab^a+a_4\,\overline{c}^a\partial _\mu A^{a\mu
}+a_5\xi \varepsilon ^{ab}\overline{c}^a\overline{c}^bc\right.  \nonumber \\
&&\,\,\,\left. +a_6\varepsilon ^{ab}\overline{c}^aA_\mu A^{b\mu
}+a_7\,c^{*}c+a_8(A^{*\mu }+\partial ^\mu \overline{c})A_\mu \right) \;.
\label{delta}
\end{eqnarray}
In the derivation of the expression for $\Sigma ^{\mathrm{c}}$ in eq.$\left( 
\ref{b-c}\right) $ use has been made of the discrete symmetry

\begin{eqnarray}
\Phi ^1 &\rightarrow &\Phi ^1  \nonumber \\
\Phi ^2 &\rightarrow &-\Phi ^2  \nonumber \\
\Phi ^{\mathrm{diag}} &\rightarrow &-\Phi ^{\mathrm{diag}}  \label{dissym}
\end{eqnarray}
where $\Phi ^a$ and $\Phi ^{\mathrm{diag}}$ stand respectively for all
off-diagonal and diagonal fields and external sources. As one can easily
recognizes, this symmetry plays the role of the charge conjugation.

Moreover, taking into account the conditions $\left( \ref{g-c}\right) $, $%
\left( \ref{u1-c}\right) $ and

\begin{equation}
\mathcal{GB}_\Sigma +\mathcal{B}_\Sigma \mathcal{G=W}^{\mathrm{U(1)}}\;,
\label{comm}
\end{equation}
one obtains

\begin{equation}
a_5=-a_3,\;\;\;a_6=-a_4,\;\;\;a_7=0,\;\;\;a_8=0\;.  \label{a}
\end{equation}
Finally, the most general invariant counterterm $\Sigma ^{\mathrm{c}}$
compatible with all Ward identities and constraints contains five free
independent parameters and reads

\begin{eqnarray}
\Sigma ^{\mathrm{c}} &=&-\frac \sigma {4g^2}\int d^4x\,\left( F_{\mu \nu
}^aF^{a\mu \nu }+F_{\mu \nu }F^{\mu \nu }\right) +\mathcal{B}_\Sigma \,\int
d^4x\,\left( a_1\,c^{a*}c^a+a_2\,A_\mu ^{a*}A^{a\mu }\right.  \nonumber \\
&&\,\,\,\left. +a_3\,\xi \overline{c}^a\left( b^a+\varepsilon ^{ba}\overline{%
c}^bc\right) +a_4\,\overline{c}^aD_\mu ^{ab}A^{b\mu }\right) \;.
\label{sigmac}
\end{eqnarray}
The free parameters $(\sigma ,\;a_{1,\;}a_2,\;a_3,\;a_4)\;$ are responsible
for the renormalization of the coupling constant $g$, of the gauge parameter 
$\xi $ and of the field amplitudes. Indeed, it is immediate to check that
the counterterm $\Sigma ^{\mathrm{c}}$ can be reabsorbed into the classical
action $\Sigma $

\begin{equation}
\Sigma +\epsilon \Sigma ^{\mathrm{c}}=\Sigma (g_0,\xi _0,\,\Phi _0^a,\Phi
_0^{\mathrm{diag}})\;+O(\epsilon ^2)\;,  \label{re}
\end{equation}
by redefining the couplings and the field amplitudes according to

\begin{eqnarray}
g_0 &=&Z_gg\,,\,\,\,\,\,\,\,\,\xi _0=Z_\xi \xi \,  \nonumber \\
A_0^a &=&Z_AA^a\,,\,\,\,A_0^{a*}=Z_A^{-1}A^{a*}  \nonumber \\
c_0^a &=&Z_cc^a\,,\,\,\,\,\,c_0^{a*}=Z_c^{-1}c^{a*}  \nonumber \\
\,\overline{c}_0^a &=&Z_{\overline{c}}\overline{c}^a\,,\,\,\,\,\,b_0^a=Z_{%
\overline{c}}b^a  \label{rcons}
\end{eqnarray}
and

\begin{equation}
\Phi _0^{\mathrm{diag}}=\Phi ^{\mathrm{diag}}\;  \label{phi-d}
\end{equation}
with

\begin{eqnarray}
Z_g &=&1+\epsilon z_g=1-\epsilon \frac \sigma 2  \nonumber \\
Z_\xi  &=&1+\epsilon z_\xi =1+2\epsilon \left( a_3-a_4\right)   \nonumber \\
Z_A &=&1+\epsilon z_A=1+\epsilon \,a_2  \nonumber \\
Z_c &=&1+\epsilon z_c=1-\epsilon \,a_1  \nonumber \\
Z_{\overline{c}} &=&1+\epsilon z_{\overline{c}}=1+\epsilon \,a_4\;.
\label{Zs}
\end{eqnarray}
It is worth noting here that the diagonal fields do not need to be
renormalized. For the diagonal ghost $c$ this property follows directly from
the ghost equation $\left( \ref{ge}\right) ,$ which implies the vanishing of
the coefficient $a_7$ in expression $\left( \ref{delta}\right) .$ For the
diagonal gauge field $A_{\mu \text{ }}$ this feature relies on the $U(1)$
Ward identity $\left( \ref{u1ident}\right) $ and on the parametrization
which has been adopted for the coupling constant $g$ in the Yang-Mills
action $\left( \ref{sym}\right) .$ We recall indeed that the parametrization
in which $g$ appears only in the interaction vertices and not in the
quadratic term is obtained by a simple redefinition of the gauge field. With
this parametrization the diagonal gauge field will acquire a nonvanishing
wave function renormalization which, however, turns out to coincide with the
inverse of the renormalization constant of the coupling $g\;$\cite
{mlp,quandt,kondo1}. Again, this feature reminds us the Ward identity of $%
QED.$ Let us also remark that the different value for the wave function
renormalization constants of the diagonal and off-diagonal components of the
gauge fields is due to the lack of the global $SU(2)$ invariance, which is
broken by the gauge fixing \cite{mlp}.

\subsection{The Callan-Symanzik equation}

Having proven the stability under radiative correction of the MAG with $%
\alpha =\xi ,$ let us turn to analyse the consequences of the diagonal ghost
Ward identity on the Callan-Symanzik equation. To this end we need to
establish a basis \cite{book} of classical insertions for the invariant
counterterm $\Sigma ^{\mathrm{c}}$ in eq.$\left( \ref{sigmac}\right) .$ This
is easily done by rewriting the counterterm in the parametric form

\begin{equation}
\Sigma ^{\mathrm{c}}=\left( z_gg\frac \partial {\partial g}+z_\xi \xi \frac
\partial {\partial \xi }+z_A\mathcal{N}_{A^a}+z_c\mathcal{N}_{c^a}+z_{%
\overline{c}}\left( \mathcal{N}_{\overline{c}^a}+\mathcal{N}_{b^a}\right)
\right) \Sigma  \label{cpar}
\end{equation}
where $\mathcal{N}_{A^a}$,$\;\mathcal{N}_{c^a}$, $\mathcal{N}_{\overline{c}%
^a}$ and $\mathcal{N}_{b^a}$ stand for the counting operators

\begin{eqnarray}
\mathcal{N}_{A^a} &=&\int d^4x\left( A_\mu ^a\frac \delta {\delta A_\mu
^a}\;-A_\mu ^{a*}\frac \delta {\delta A_\mu ^{a*}}\right) \;,  \nonumber
\label{count} \\
\mathcal{N}_{c^a} &=&\int d^4x\left( c^a\frac \delta {\delta
c^a}\;-c^{a*}\frac \delta {\delta c^{a*}}\right) \;,  \nonumber \\
\mathcal{N}_{\overline{c}^a} &=&\int d^4x\;\overline{c}^a\frac \delta
{\delta \overline{c}^a}\;\;,\;\;\;\;\;\;\;\mathcal{N}_{b^a}=\int
d^4x\;b^a\frac \delta {\delta b^a}\;\;.  \label{coper}
\end{eqnarray}
Therefore, for the Callan-Symanzik equation we get \cite{book}

\begin{equation}
\mu \frac{\partial \Gamma }{\partial \mu }+\beta _g\frac{\partial \Gamma }{%
\partial g}+\beta _\xi \frac{\partial \Gamma }{\partial \xi }-\gamma _{A^a}%
\mathcal{N}_{A^a}\Gamma -\gamma _{c^a}\mathcal{N}_{c^a}\Gamma -\gamma _{%
\overline{c}^a}\left( \mathcal{N}_{\overline{c}^a}+\mathcal{N}_{b^a}\right)
\Gamma =0\;,  \label{cso}
\end{equation}
where $\Gamma =\Sigma +O(\hbar )$ is the vertex functional and $\mu $ the
renormalization point.

We see therefore that, as a consequence of the diagonal ghost equation and
of the $U(1)$ Ward identity, the anomalous dimensions of $c$ and $A_\mu \;$%
vanish. Also, acting on $\left( \ref{cso}\right) $ with the test operator $%
\delta ^2/\delta A_\mu \delta A_\nu \;$and setting to zero all fields and
the gauge parameter $\xi $, it is apparent that the beta function $\beta _g$
of the gauge coupling can be obtained directly from the two-point Green
function with diagonal gauge fields as external legs \cite{quandt}.

\section{Generalization to the gauge group SU(N)}

In this section we shall discuss the generalization of the diagonal ghost
equation to the case of the gauge group $SU(N).$ We shall limit ourselves to
the classical aspects since the quantum extension can be straightforwardly
achieved by repeating the same procedure as before. Let us begin by
reminding that for $SU(N)$ the Cartan subgroup has $N-1$ generators.
Therefore, adopting the MAG condition, the original gauge group is broken
down to $U(1)^{N-1}$. Accordingly, the gauge connection decomposes as follows

\begin{equation}
\mathcal{A}_\mu =\mathcal{A}_\mu ^AT^A=A_\mu ^aT^a+A_\mu ^iT^{\,i},
\label{conn-sun}
\end{equation}
where the index $i$ labels the generators $T^{\,i}$ of the Cartan subgroup
and runs from $1$ to $N-1.\;$The remaining off-diagonal generators $T^a$
will be labelled by the index $a$ running from $1$ to $N(N-1).\;$In
particular, from the Jacobi identity for the structure constants $f^{ABC%
\text{ }}$ of $SU(N)$

\[
f^{ABC\text{ }}f^{CMN\text{ }}+f^{AMC\text{ }}f^{CNB\text{ }}+f^{ANC\text{ }%
}f^{CBM\text{ }}=0\;, 
\]
one easily obtains the following identities 
\begin{eqnarray}
f\,^{abi}f^{\,bjc}+f\,^{abj}f^{\,bci} &=&0\;,\;\;  \nonumber  \label{jacn} \\
f\,^{abc}f^{\,bdi}+f\,^{abd}f^{\,bic}+f\,^{abi}f^{\,bcd} &=&0\;,
\label{jacn}
\end{eqnarray}
where $\left( i,j=1,...,N-1\right) $ and $\left( a,b,c,d=1,...,N(N-1)\right)
.$ These identities will turn out to be very useful for the generalization
of the diagonal ghost equation.

In the present case, the BRST transformations read

\begin{eqnarray}
sA_\mu ^a &=&-\left( D_\mu ^{ab}c^b+f^{\,abc}A_\mu ^bc^c+f^{\,abi}A_\mu
^bc^i\right) ,\,\,\,\;sA_\mu ^i=-\left( \partial _\mu c^i+f\,^{iab}A_\mu
^ac^b\right) \;,  \nonumber \\
sc^a &=&f\,^{abi}c^bc^i+\frac
12f\,^{abc}c^bc^c,\,\,\,\,\,\,\,\,\,\,\,\,\,\,\,\,\,\,\,\,\,\,\,\,\,\,\,\,\,%
\,\,\,\,\,\,\,\,\,\,\,sc^i=\frac 12\,f\,^{iab}c^ac^b,  \nonumber \\
s\overline{c}^a
&=&b^a\;,\,\,\,\,\,\,\,\,\,\,\,\,\,\,\,\,\,\,\,\,\,\,\,\,\,\,\,\,\,\,\,\,\,%
\,\,\,\,\,\,\,\,\,\,\,\,\,\,\,\,\,\,\,\,\,\,\,\,\,\,\,\,\,\,\,\,\,\,\,\,\,\,%
\,\,\,\,\,\,\,\,\,\,\,\,\,\,\,s\overline{c}^i=b^i\;,  \nonumber \\
sb^a
&=&0\;,\,\,\,\,\,\,\,\,\,\,\,\,\,\,\,\,\,\,\,\,\,\,\,\,\,\,\,\,\,\,\,\,\,\,%
\,\,\,\,\,\,\,\,\,\,\,\,\,\,\,\,\,\,\,\,\,\,\,\,\,\,\,\,\,\,\,\,\,\,\,\,\,\,%
\,\,\,\,\,\,\,\,\,\,\,\,\,\,\,\,\,\,sb^i=0\;.  \label{BRSTn}
\end{eqnarray}
For the decomposition of the field strength we get

\begin{equation}
\mathcal{F}_{\mu \nu }=\mathcal{F}_{\mu \nu }^AT^A=F_{\mu \nu }^aT^a+F_{\mu
\nu }^iT^{\,i}\;,  \label{fsn}
\end{equation}
with the off-diagonal and diagonal parts given respectively by

\begin{eqnarray}
F_{\mu \nu }^a &=&D_\mu ^{ab}A_\nu ^b-D_\nu ^{ab}A_\mu ^b\;+\,f^{abc}A_\mu
^bA_\nu ^c\;,  \nonumber \\
F_{\mu \nu }^i &=&\partial _\mu A_\nu ^i-\partial _\nu A_\mu ^i+f^{abi}A_\mu
^aA_\nu ^b\;\,,  \label{fscompn}
\end{eqnarray}
where the covariant derivative $D_\mu ^{ab}$ is defined with respect to the
diagonal components $A_\mu ^i$

\begin{equation}
D_\mu ^{ab}\equiv \partial _\mu \delta ^{ab}-f^{abi}A_\mu ^i\,\,\,\,\,\,.
\label{cdern}
\end{equation}
The Yang-Mills action is now found to be 
\begin{equation}
S_{\mathrm{YM}}=-\frac 1{4g^2}\int d^4x\,\left( F_{\mu \nu }^aF^{a\mu \nu
}+F_{\mu \nu }^iF^{i\mu \nu }\right) \;.  \label{symn}
\end{equation}
Also, the MAG\ and the diagonal gauge-fixing terms are generalized to

\begin{equation}
S_{\mathrm{MAG}}=s\,\int d^4x\,\left( \overline{c}^a\left( D_\mu
^{ab}A^{b\mu }+\frac \xi 2b^a\right) -\frac \xi 2f\,^{abi}\overline{c}^a%
\overline{c}^bc^i-\frac \xi 4f\,^{abc}c^a\overline{c}^b\overline{c}^c\right)
\;,  \label{smn}
\end{equation}

\begin{equation}
S_{\mathrm{diag}}=s\,\,\int d^4x\,\,\overline{c}^i\partial ^\mu A_\mu ^i\;.
\label{su1n}
\end{equation}
Notice that the MAG\ condition in eq.$\left( \ref{smn}\right) $ contains the
unique gauge parameter $\xi .\;$ To write down the Slavnov-Taylor identity
we introduce external sources

\begin{equation}
S_{\mathrm{ext}}=\int d^4x\left( A_\mu ^{a*}sA^{a\mu }+A_\mu ^{i*}sA^{i\mu
}+c^{a*}sc^a+c^{i*}sc^i\right) \;,  \label{sextn}
\end{equation}
so that the complete action

\begin{equation}
\Sigma =S_{\mathrm{YM}}+S_{\mathrm{MAG}}+S_{\mathrm{diag}}+S_{\mathrm{ext}}
\label{ca}
\end{equation}
obeys the Slavnov-Taylor identity

\begin{equation}
\mathcal{S}(\Sigma )=0\;,  \label{stn}
\end{equation}
with

\begin{equation}
\mathcal{S}(\Sigma )=\int d^4x\left( \frac{\delta \Sigma }{\delta A_\mu ^{a*}%
}\frac{\delta \Sigma }{\delta A^{a\mu }}+\frac{\delta \Sigma }{\delta A_\mu
^{i*}}\frac{\delta \Sigma }{\delta A^{i\mu }}+\frac{\delta \Sigma }{\delta
c^{a*}}\frac{\delta \Sigma }{\delta c^a}+\frac{\delta \Sigma }{\delta c^{i*}}%
\frac{\delta \Sigma }{\delta c^i}+b^a\frac{\delta \Sigma }{\delta \overline{c%
}^a}+b^i\frac{\delta \Sigma }{\delta \overline{c}^i}\right) \;
\label{stn-exp}
\end{equation}
In order to derive the diagonal ghost equation we proceed as before and we
apply the functional operator 
\begin{equation}
\mathcal{G}^i\mathcal{=}\frac \delta {\delta c^i}+f\,^{abi}\overline{c}%
^a\frac \delta {\delta b^b}\;,  \label{gopn}
\end{equation}
to the complete action $\Sigma .$ Using then the Jacobi identities $\left( 
\ref{jacn}\right) ,$ we obtain

\begin{equation}
\mathcal{G}^i\Sigma =\Delta _{\mathrm{cl}}^i\;,  \label{gen}
\end{equation}
where

\begin{equation}
\Delta _{\mathrm{cl}}^i=-\partial ^2\,\overline{c}^i+f\,^{abi}A_\mu
^{a*}A^{b\mu }-\partial ^\mu A_\mu ^{i*}-f\,^{abi}c^{a*}c^b\;.  \label{cbn}
\end{equation}
Equation $\ \left( \ref{gen}\right) $ is the generalization of the diagonal
ghost equation we are looking for. Again, the breaking term $\left( \ref{cbn}%
\right) $ is purely linear in the quantum fields and will  not be affected
by radiative corrections.

It is worth underlining that the diagonal Ward identity corresponding to the 
$U(1)^{N-1\text{ }}$Cartan subgroup follows from anticommuting the ghost
equation $\left( \ref{gen}\right) $ with the Slavnov-Taylor identity $\left( 
\ref{stn}\right) ,$ \textit{i.e.} 
\begin{equation}
\mathcal{G}^i\mathcal{S}(\Sigma )=0\,\,\,\,\,\,\,\Rightarrow \;\;\mathcal{W}%
^i\Sigma =-\partial ^2b^i\;,  \label{gh-u1}
\end{equation}
where $\mathcal{W}^i\;$is the Ward operator

\begin{equation}
\mathcal{W}^i=\partial _\mu \frac \delta {\delta A_\mu ^i}+f^{abi}\left(
A_\mu ^a\frac \delta {\delta A_\mu ^b}+c^a\frac \delta {\delta c^b}+b^a\frac
\delta {\delta b^b}+\overline{c}^a\frac \delta {\delta \overline{c}^b}+A_\mu
^{a*}\frac \delta {\delta A_\mu ^{b*}}+c^{a*}\frac \delta {\delta
c^{b*}}\right) \;.  \label{wop}
\end{equation}
of the residual $U(1)^{N-1\text{ }}$subgroup. As in the case of $SU(2),$ the
diagonal components $A_\mu ^i$ behave as photons, while all off-diagonal
fields play the role of charged matter.

Let us end up this section by remarking that the whole set of Ward
identities is easily extended to the quantum level. As before, they imply
the vanishing of the anomalous dimensions of the diagonal ghost $c^i$ and of 
$A_\mu ^i$.

\section{Conclusions}

We have presented a BRST algebraic analysis of $SU(N)$ Yang-Mills theories
in a class of maximal Abelian gauges with only one parameter. The existence
of a new Ward identity, called the diagonal ghost equation, has been pointed
out. It is a nonintegrated linearly broken identity which plays a crucial
role in the proof of the stability of the theory under radiative corrections
and which turns out to have many consequences. For instance, the Abelian
Ward identity corresponding to the Cartan subgroup follows by applying the
diagonal ghost equation to the Slavnov-Taylor identity, displaying therefore
a close similarity with $QED$.

The diagonal ghost equation provides strong constraints on the
Callan-Symanzik equation, implying the vanishing of the anomalous dimension
of the diagonal ghosts. In addition, a simple proof of the fact that the
beta function of the gauge coupling can be obtained directly from the vacuum
polarization with diagonal gauge fields as external legs has been given. To
some extent, these results might be interpreted as further evidences for the
infrared Abelian dominance in the MAG.

Let us conclude by noting that the ghost equation Ward identity could
provide a strong indication of the decoupling of the diagonal ghosts at low
energy. This important point should also be related to the ghost
condensation mechanism proposed in \cite{kondo,schaden}$.$ Suppose indeed
that all off-diagonal fields become massive, due to the existence of a
condensation mechanism taking place at some energy scale, which sets the
confinement scale. Therefore, below this scale the off-diagonal components
decouple. Concerning now the diagonal ghost equation, it is not difficult to
see that it implies that the vertex functional $\Gamma $ depends on $c^i$
and the off-diagonal Lagrange multipliers $b^a$ through the combination $%
\widehat{b}^a=(b^a-f^{abi}\overline{c}^bc^i).$ Then, performing in the path
integral the change of variables $b^a\rightarrow $ $\widehat{b}^a$, $\,%
\overline{c}^a\rightarrow \overline{c}^a$,\thinspace $\,c^i\rightarrow c^i$,
whose Jacobian is equal to one, it turns out that the diagonal ghosts $c^i$
contribute to the transformed action through the bilinear term $\overline{c}%
^i\partial ^2c^i,$up to terms in the external classical sources. It is only
the diagonal antighosts $\overline{c}^i$ which appear in the interaction
terms. However, it is remarkable that $\overline{c}^i$ interact only with
the off-diagonal fields. These interaction terms should therefore be
suppressed below the confinement scale due to the condensates of the
off-diagonal fields, implying thus the decoupling of the diagonal ghosts.
This aspect is actually under investigation and will be reported in a
forthcoming work.

\section*{Acknowledgements}

A.R. Fazio thanks M. Schaden and M. Picariello for fruitful discussion and
the Theoretical Physics Department of the State University of Rio de Janeiro
(UERJ)\ for kind hospitality. The Conselho Nacional de Desenvolvimento
Cient\'{\i}fico e Tecnol\'{o}gico CNPq-Brazil, the Funda{\c {c}}{\~{a}}o de
Amparo {\`{a}} Pesquisa do Estado do Rio de Janeiro (Faperj), the SR2-UERJ,
the INFN (sezione di Milano) and the MURST, Ministero dell' Universit\'{a} e
della Ricerca Scientifica e Tecnologica, Italy, are acknowledged for the
financial support.

\end{document}